*Darwinian Selection Induces Lamarckian Adaptation in a Holobiont Model*

Osmanovic D[1], Kessler DA[1], Rabin Y[1,2], Soen Y[3*]

[1] Department of Physics, Bar-Ilan University, Ramat Gan 52900, Israel

[2] NYU-ECNU Institute of Physics at NYU Shanghai 200062, China

[3] Department of Biomolecular Sciences, Weizmann Institute of Science, Rehovot 76100, Israel

[*] Corresponding author. Email: yoavs@weizmann.ac.il

**Abstract**

Current models of animal evolution focus on selection of individuals, ignoring the much faster selection of symbiotic bacteria. Here we take host-symbiont interactions into account by introducing a Population Genetics-like model of holobionts exposed to toxic stress. The stress can be alleviated by selection of resistant individuals (host and bacteria) and by secretion of a detoxification agent ("detox"). By defining a new measure, termed the '*Lamarckian*', we show that selection of resistant bacteria over one generation of hosts leads to stress-dependent increase in the tolerance of the hosts' offspring. This benefit is mediated by co-alleviation of toxic and physiologic stress. Prolonged exposure leads to further adaptation by 'group selection' of bacterial communities with higher detox per bacterium. These findings show that Lamarckian adaptation can arise via interactions between two levels of Darwinian selection within a holobiont system. The conclusions and modelling framework are applicable to diverse types of holobiont systems.

**Introduction**

Evolutionary adaptations are commonly thought to be driven by genetic mutations occurring on a timescale of many generations. Selection of individuals with rare beneficial mutations can then enable the adaptation of a large population without the emergence of new variations within a single lifetime. This view has recently been brought into question (*1-7*) by evidence for non-Mendelian epigenetic phenomena (*8-11*), genome editing and mobility systems (*12, 13*), niche construction (*14*) and contribution of symbiotic micro-organisms to heritable variation of the whole unit (holobiont) (*3, 5, 6, 15, 16*). The latter may be of particular interest because it violates the fundamental Neo-Darwinian assumption of a single level of selection per individual. Considering the individual as an interacting community with multiple levels of selection and generation times may have transformative implications on the dynamics and outcomes of evolution (*3, 5, 6*). The genetic content and species composition of symbiotic bacteria can change substantially within a single generation of the host and are inherited by vertical and/or horizontal transmission of bacteria (*3, 5, 6, 17-19*). Since the symbiotic microbiome is an integral part of host development and physiology (*16*), variations in the microbiome can influence the state and phenotypes of the host. The resulting changes are not necessarily limited to somatic tissue of the host. They can also extend to the host germline (*20*) and can contribute to non-Mendelian inheritance of environmentally-induced phenotypes (*17*). In line with these considerations, it has been suggested that rapid diversification and transmission of symbiotic bacteria may contribute to rapid evolution of their host (*3, 5, 6, 15, 21*). This possibility, however, has not yet been functionally confirmed by direct experimentation and the modes of adaptation that are available for a population of holobiont communities have not been investigated in a mathematical framework. Current models of evolutionary change are typically limited to population genetics with a single species of individuals (*22*) or to ecological (*23-25*) and evolutionary game theory models (*26*) (*27*) of interactions between free-living or symbiotic (*28, 29*) species. However, these models do not consider the evolution of a holobiont population in which every individual is an interacting community of host and bacteria.

Here, we construct a general modeling framework of host-symbiont evolution under exposure to toxic stress. The holobiont in this 'Population Symbio-Genetics' (PSG) model is considered as a single unit undergoing two levels of Darwinian selection occurring on different timescales for host and bacteria. We show that this structure can exhibit Lamarckian offspring adaptation due to stress-induced during the lifetime of the parental holobiont. The Lamarckian adaptation is enabled by selection and transfer of resistant bacteria and is mediated by context-dependent alleviations of distinct types of stress (toxic and physiologic). Beyond this Lamarckian effect we show that Darwinian selection over timescales larger than a host generation, promote 'group selection' of bacterial communities with higher average detoxification per bacterium (in contrast to selection of better fit individual bacteria which takes place within a host generation). This form of group selection enhances the adaptation independently of the Lamarckian effect and is accompanied by stress-dependent variability across holobionts (despite the fixed mutation rate).

**Formulation of the PSG Model**

The model considers a population of holobionts under exposure to active toxin of concentration $T$. Interactions between each host and its bacterial symbionts are mediated by: (i) mutual alleviation of toxic challenge via secretion of a detoxification agent ("detox"), (ii) dependence of the hosts' well-being on the size of the bacterial population and (iii) modulation of the bacterial niche size by the stress state of the host.

For each host and bacterium, we define the probability of survival to reproduction ($P_H$ and $P_B$, respectively), as follows:

(1)  $P_H = (1 - N_H/2K_H) \exp[-(\hat{S}_H + \hat{S}_{Ph})]$

(2)  $P_B = (1 - N_B/2K_B) \exp(-S_B)$

Here, $N_H$ is the number of hosts and $K_H$ is the maximal number of hosts that can be supported by the external environment (carrying capacity for hosts). $K_B$ is the number of bacteria that can be accommodated in the host (carrying capacity for bacteria) and $N_B$ is the number of bacteria per host. Representing host- and bacterial-specific sensitivities to toxin by $x_H$ and $x_B$, respectively, we define the instantaneous toxic stress to host and bacteria as $S_H = x_H T$ and $S_B = x_B T$. $\hat{S}_H = <S_H>_t$ represents the average of $S_H$ over a host generation time (interval between host reproduction events). Additionally, $\hat{S}_{Ph}(N_B) = ln(<N_B>_t/K_B^0) + (1 - <N_B>_t/K_B^0)$ corresponds to a 'physiological stress' over a host generation due to deviations from the bacterial population size that is preferred in toxin-free conditions. This size is determined by $K_B^0$, which also provides an inverse scale ($1/K_B^0$) for the negative impacts of losing bacterial-derived metabolites (when $N_B < K_B^0$) or having to support excess numbers of bacteria ($N_B > K_B^0$)(30). Accordingly, the physiological stress vanishes when $N_B = K_B^0$. The averages over a host generation (i.e. $<S_H>_t$ and $<N_B>_t$) are introduced as practical approximations, enabling us to simplify the model by evaluating the survival of the host only at the time of reproduction.

Modulation of the bacterial niche by the state of the host (*31-34*) is modelled by changing $K_B$ as a function of the instantaneous host stress:

(3)  $K_B = K_B^0 (1 + \delta S_H)$

where $\delta$ is a host-intrinsic property, determining the direction and extent of the change in the bacterial niche size due to the stress of the host. Since bacteria can affect this stress by secreting detox on a timescale shorter than a host generation, $K_B$ is jointly influenced by the host and the bacteria. To enable an unbiased analysis of how $K_B$ might evolve under different exposures to toxin, we consider a starting population of hosts with an broad distribution of $\delta$'s, symmetric around zero.

We assume that all the hosts and their bacteria are exposed, at time *t*, to the same influx of active toxin, $\vartheta(t)$, applied instantaneously (i.e., in one bacterial generation, $\Delta t$). This toxin can be neutralized in a holobiont-specific manner by release of detox from the host and each of its bacteria (*30, 34-36*):

(4)  $T(t+\Delta t) = T(t) \exp(-\lambda_B \sum y_B - \lambda_H y_H) + \vartheta(t)$

where $y_H$ and $y_B$ are the amounts of detox secreted inside a given holobiont at time *t*, and $\lambda_H$ and $\lambda_B$ are constant detoxification capacities. While all the individuals within a given holobiont equally benefit from the total amount of secreted detox, they have individual-specific sensitivities to toxin.

The evolving traits of the host and its bacteria (*x, y* and $\delta$) are initially drawn from trait-specific distributions. Surviving bacteria divide at every time step of the simulation ($\Delta t$) and living hosts reproduce every $\tau$ generations of bacteria (so that host generation time is $\tau \Delta t$). We consider the simplest reproduction model in which each of the surviving hosts and bacteria gives rise to one offspring that inherits the traits of its parent, subject to a small random modification depending on a constant mutation rate, $\mu$ (Supplementary Methods).

(5)  $z_{\text{offspring}} = z_{\text{parent}} + \eta \sqrt{\mu} - \beta_z \mu (z_{\text{parent}} - z_0)$

Here *z* corresponds to any of the evolving traits *x, y* and $\delta$, $\eta$ is a standard Gaussian deviate with zero mean, and $z_0$ and $\beta_z$ are trait-specific coefficients, controlling the peak and width of the steady state distributions (specified in Supplementary information). Values of $\beta_y$ and $\beta_\delta$ were chosen to support wide distributions of *y* and $\delta$, respectively. For the sensitivity traits ($z = x_H$ and $x_B$), the distribution is truncated at *x* = 0, so as to prevent a trivial solution in which all the individuals are completely insensitive to toxin. We avoid negative values of detox secretion by setting negative *y* values in Eq. 5 to zero. The remaining dynamic variables are updated in every generation of bacteria ($N_B$, *T*, $S_H$, $S_B$ and $K_B$) and host ($N_H$). In this study, we consider an initial population of 32000 hosts ($N_H = K_H$ = 32000) with 100 bacteria per host ($N_B = K_B^0$ = 100). We set the host generation time to $\tau$ = 100 bacterial generations and all mutation rates to $\mu$ = $10^{-3}$ per generation (for both host and bacteria).

**Stress-dependent adjustment of bacterial niche size**

We first examined the effects of exposure to a single pulse of toxin, $T_0$, applied at $t_0$ (i.e. $\vartheta(t_0)=T_0$). On timescales smaller than one host generation (100$\Delta t$), the bacterial community undergoes selection for less sensitive bacteria, accompanied by a drop in the bacterial population (Fig. 1A,B). In a system with only one level of selection (e.g. free-living bacteria), this would be the only adaptive change. However, when the bacterial population is coupled to a live host, the survival of the holobiont depends also on the amount of detox secreted by the bacterial community (Fig. 1C). This secretion is higher for hosts which react to the toxic stress by increasing their carrying capacity for bacteria (i.e. hosts with $\delta > 0$; Supplementary Fig. S1A). This leads to stress-dependent selection of hosts which provide a larger bacterial niche, $K_B$ (Fig. 1D) and increases the average size of the recovered bacteria population beyond $K_B^0$ (Fig. 1A). The benefit from larger $K_B$ is two-fold: It reduces the negative impact of losing bacteria (by assisting recovery of the bacterial population; Fig. 1E, Supplementary Fig. S1B) and increases the total amount of secreted detox (Fig. 1F, Supplementary Fig. S1C). However, when $N_B > K_B^0$ the benefit from higher detox secretion is accompanied by the negative impact of bacterial overload. The combination of the two opposing effects adjusts the size of the bacterial population in a stress-dependent manner which maximizes the probability of survival of the holobiont as a whole.

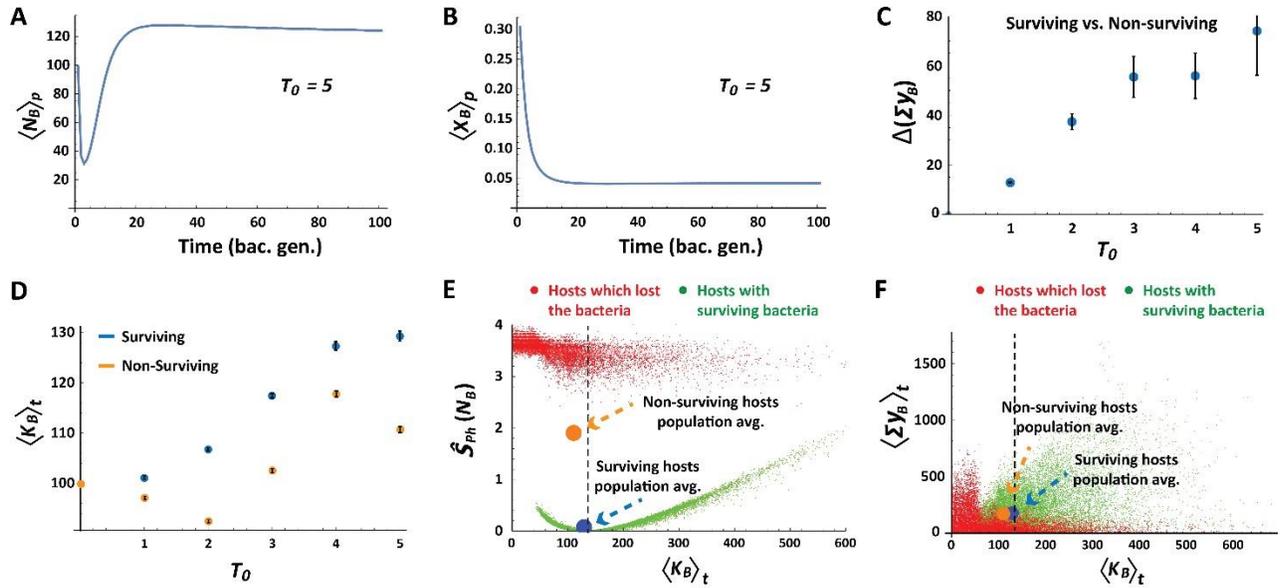

**Figure 1: Stress-dependent adjustment of the average bacterial niche size**. **(A,B)** Short term kinetics of the population averaged number of bacterial symbionts, $<N_B>_P$ (A) and bacterial sensitivity, $<X_B>_P$ (B) for hosts which survived a single pulse of exposure to toxin, $T_0$=5, applied at the initial time step. **(C)** Average difference $\pm$ standard error (SE) between surviving and non-surviving hosts with respect to the total amount of detox secreted by bacteria over the entire host generation (shown for each $T_0$). **(D)** Mean carrying capacity for bacteria in the population of hosts, averaged ($\pm$ SE) over one host generation at different levels of $T_0$. **(E)** The physiological stress over a host generation, $\hat{S}_{Ph}(N_B)$, versus time average of its carrying capacity for bacteria. Green and red points represent hosts with surviving and non-surviving bacteria, respectively. Blue and orange circles mark population averages for surviving and non-surviving holobionts, respectively. Dotted line marks carrying capacity which minimizes the physiological stress. **(F)** Same as (E) for the time average of total bacterial detox versus bacterial carrying capacity. Time and population averages are denoted by $t$ and $p$ subscripts, respectively).

**Stress-dependent Lamarckian adaptation**

Bacteria that are modified by the stress in one host generation can be transmitted to the next generation, potentially increasing the stress tolerance of the offspring. To quantitatively evaluate this possibility we introduce a new measure, termed the "*Lamarckian*", representing the gain in offspring tolerance due to acquisition of traits by its parent. To evaluate the contribution of acquired (as opposed to initial) traits, we identify the subpopulation of hosts that survived the first generation of exposure and apply a new simulation to these parents at their initial state ("cloned parents") and to their offspring (Fig. 2A). We then compare the survival rates of the offspring ($SR_{Offs}$) to that of their "cloned parents" ($SR_{CP}$). The use of a parental subpopulation with the original state of microbiota allows us to distinguish increase of tolerance due to selection of initially better fit parents from a gain of tolerance due to transmission of changes acquired during a host generation (not present in the initial parental clones). We then define the *Lamarckian*, **L**, as:

(6)   **L** = $SR_{Offs} / SR_{CP} - 1$,

so that it is positive if the average tolerance increases due to transfer of changes acquired during a host generation.

For a given $\lambda_B$, we find that **L** is an increasing function of the injected amount of toxin and vanishes at low $T_0$ (Fig. 2B). Analysis of **L** for different choices of $\lambda_B$ (at a given $T_0$) reveals a non-monotonic dependence on $\lambda_B$, manifested by an essentially constant **L** > 0 over a range of small $\lambda_B$, followed by an increase to a maximum at intermediate values of $\lambda_B$ and lastly, a decline at sufficiently large $\lambda_B$ (Fig. 2C). The positive *Lamarckian* is the result of transferring bacterial population which acquired toxin resistance during the parental host generation (Fig. 2D). To determine how this bacterial transfer confers a gain of toxin tolerance in the hosts' offspring, we compared the offspring and their cloned parents with respect to the toxic and physiologic stress under exposure to the toxin. For small enough $\lambda_B$, the reduction of toxic stress by bacteria is negligible and the positive *Lamarckian* is primarily due to alleviation of the physiological stress in the offspring (Supplementary Fig. S2A). At intermediate values of $\lambda_B$, the alleviation of bacterial loss enhances the neutralization of toxin, providing an additional contribution to the Lamarckian (Fig. 2E,F). However, when $\lambda_B$ is large enough to support substantial neutralization of toxin during a single host generation (Supplementary Fig. S3), the Lamarckian decreases due to the reduction of stress in both the parents and the offspring (Supplementary Fig. S2B).

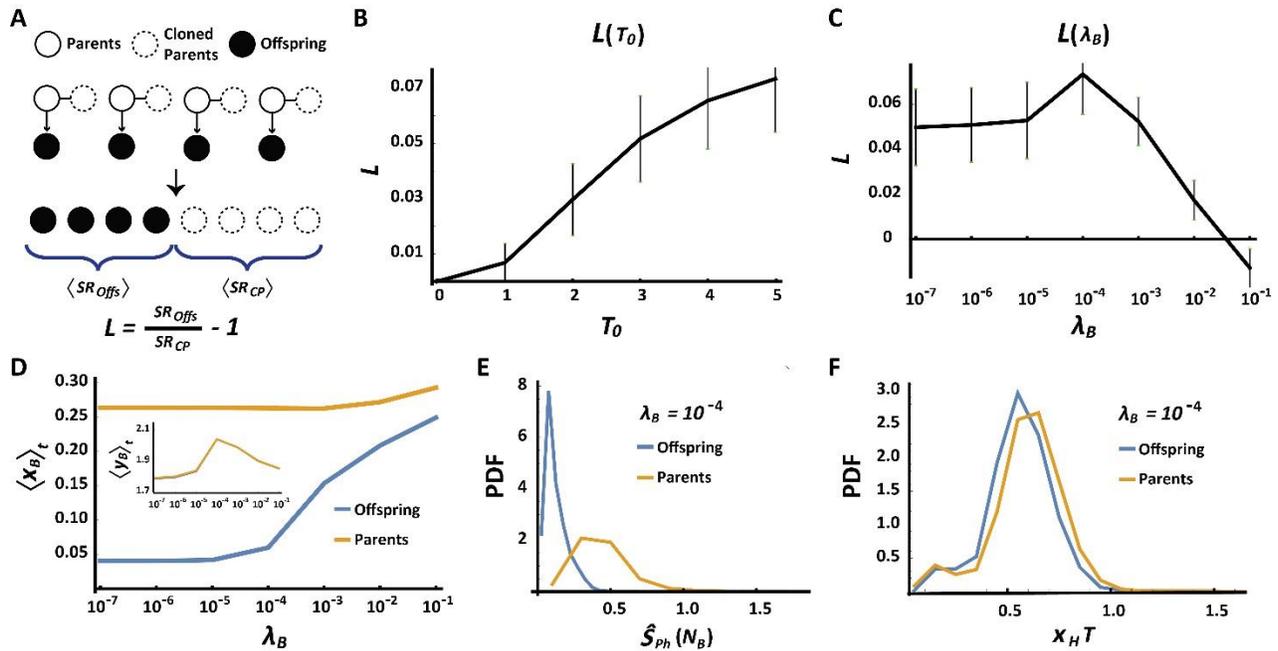

**Figure 2: Stress-dependent Lamarckian adaptation. (A)** Schematics of the Lamarckian evaluation protocol. **(B, C)** The *Lamarckian* as a function of toxic exposure (B) and bacterial detox coefficient (C). **(D)** Bacterial sensitivity and detox per bacteria (inset) as a function of bacterial detox coefficient, after exposure to toxin ($T_0$=5). Shown are time (and population) averages over one generation of unexposed 'clones' of surviving parents (orange) and their offspring (blue). **(E,F)** Distributions of the physiologic (E) and toxic stress (F) in the cloned parents and their offspring, immediately after one generation of exposure to toxin ($T_0$=5).

**Stress-dependent group selection and 'Bacterial Assimilation'**

When the toxic pressure persists over timescales larger than one host generation (Fig. 3A), the selection favors holobionts whose bacterial communities have higher average levels of detox per bacterium, $<y_B>$ (Fig. 3B). This 'group selection' is based on a collective property of the bacterial community (as opposed to selection of individual bacteria). In the current model, this selection occurs only at the time of host reproduction. If the toxin persists over a period longer than $\mu^{-1}$ bacterial generations and the elimination of mutations is sufficiently slow (small enough $\beta_y$), the selection is accompanied by significant accumulation of bacterial mutations. This enhances the group selection for higher $<y_B>$, thus leading to toxin-dependent increase in detoxification rate (Fig. 3A, inset) and expedited holobiont adaptation (Fig. 3C). Group selection for higher detox is also accompanied by extended persistence of high detox levels (Fig. 3B) and by elevated detox variability across holobionts (Fig. 3B, inset). Additional increase of variability under stress is observed in the carrying capacity for bacteria and in the size of the bacterial population (Supplementary Fig. S4A,B).

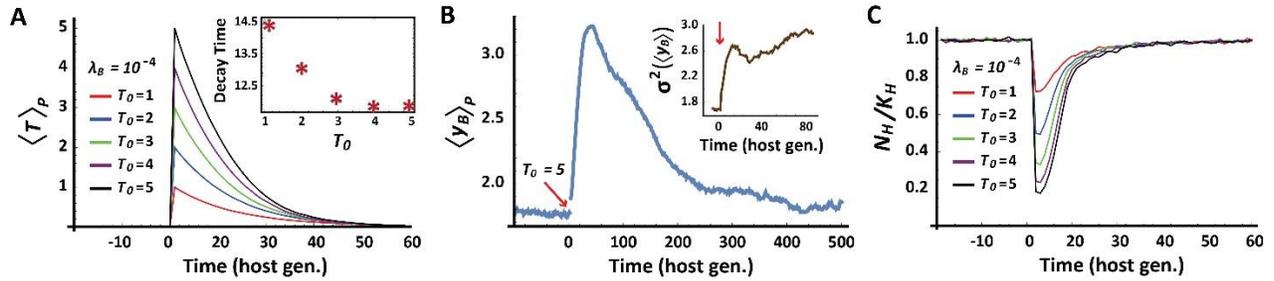

**Figure 3: Stress-dependent group selection**. **(A)** Temporal kinetics of active toxin for different initial levels of toxin, $T_0$. Inset displays the time to neutralize 50% of the toxin. **(B)** Temporal kinetics of average detox secretion per bacteria following exposure to toxin at $T_0$=5 (red arrow). $\lambda_B$=$10^{-4}$. Inset reveals an increase of inter-hosts variance in average detox per bacteria. **(C)** Kinetics of host population size, $N_H$, normalized by the host carrying capacity, $K_H$.

Bacterial mutations that were selected under stress persist over a characteristic timescale of $1/\mu$ = 10 host generations after neutralization of the toxin, thus providing 'memory' of the previous exposure. To evaluate the influence of this 'memory' on tolerance to new exposures, we analyzed the response to repeated pulses of injected toxin, separated by time intervals shorter than 10 host generations. The re-exposures lead to successive selections which oppose the relaxation of <$y_B$> to its (lower) equilibrium value (Fig. 4A vs. Fig. 3B), resulting in enhanced detoxification (Fig. 4B) and reduced impact on the host population (Fig. 4C). This enhancement reduces the environmental pressure and enables the survival of intrinsically less resistant hosts and bacteria (Fig. 4D). The progressive increase in host sensitivity due to successive selections for higher detox per bacteria, <$y_B$>, demonstrates a hitherto unrealized mode of assimilation. This mode of '*Bacterial Assimilation*' is analogous to '*Genetic Assimilation*' due to successive selections of host-intrinsic alleles (*37, 38*). However, assimilation by bacteria might be more effective because of the faster generation of a larger reservoir of variations (variations are generated by many bacteria per host on time scales that are much shorter compared to generation time of new genetic alleles in the host). This advantage of bacterial assimilation is particularly critical when the host population is small, or alternatively, when the repertoire of host-intrinsic alleles has been trimmed by previous selections.

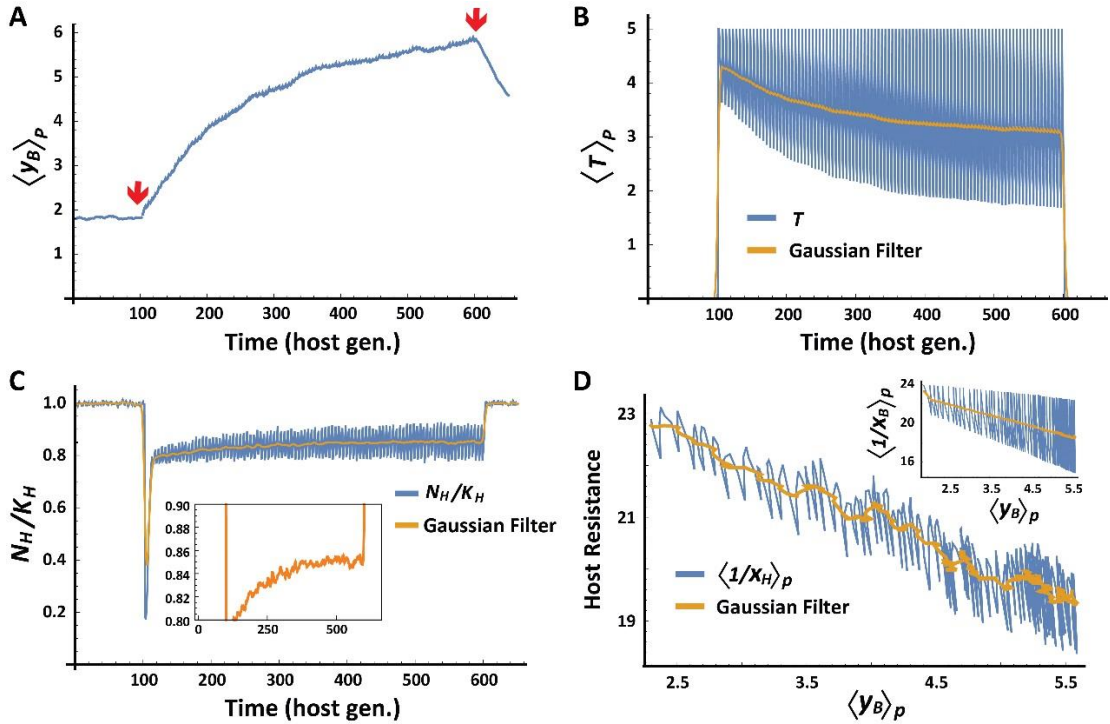

**Figure 4: Improved adaptation in response to successive exposures**. Analysis of long-term adaptation under successive resetting of the active toxin to $T$ = 5, every 5 host generations. **(A-C)** Temporal kinetic profiles of average detox per bacteria (A), active toxin (B) and normalized size of the host population (C with a magnified scale in the inset). Red arrows in (A) mark the start and end of the successive resetting if the toxin. **(D)** Inverse correlation between the increase in detox secretion per bacteria and the average toxin resistance of host, $1/x_H$, and bacteria, $1/x_B$ (inset). Orange overlays correspond to Gaussian filtering of the measured properties.

## Discussion

This work analyzes short- and long-term adaptation of a model system in which Darwinian selection of the whole (holobiont) unit is coupled to a faster Darwinian selection within the generation time of the host. It is generally accepted that every animal and plant conforms to this pattern of internal selection of microbial symbionts followed by selection of the whole organism. However, the general implications of this selection hierarchy have not yet been studied in population genetics frameworks. The latter are largely based on the Neo-Darwinian assumptions of a single level of selection per individual and rare emergences of heritable changes within the individual's lifetime (due to small mutation rate). Both assumptions are violated by considering the holobiont as a single unit. This unit is not equivalent to an assembly of independent individuals because the selection of micro-organisms within the holobiont is coupled to the selection of the holobiont as a whole. By considering the holobiont as a single unit with heritable bacteria, we show that stress-dependent Lamarckian adaptation can be induced by pure Darwinian selection without having to introduce a stress-induced change in the rate of mutations. Within the simplified model wherein the survival of the host is evaluated only at the time of reproduction, Lamarckian adaptation arises as the consequence of rapid selection and transmission of resistant bacteria. The inherited bacteria confer Lamarckian adaptation via distinct effects (alleviation of toxic stress and reduction of negative impact of bacterial loss) whose relative contributions depend on the amount of toxin exposure and detoxification capacity. Selection of holobionts with higher detox per bacteria, on the other hand, occurs on a timescale larger than one host generation and therefore cannot contribute to the *Lamarckian* which measures the offspring's gain in tolerance due to acquisition of traits in a single generation of parents. The group selection is nonetheless the main factor responsible for the progressive increase in tolerance over multiple host generations. Taken together, the Lamarckian is mediated by selection of resistant bacteria within one host generation while the longer-term adaptation under prolonged toxic pressure is achieved by group selections for higher detox per bacteria.

The absence of faster group selection reflects a lack of mechanism (in our model) for changing <$y_B$> during a single host generation (with the possible exception of rare cases of rapid changes in <$y_B$> due to amplification of very small numbers of resistant bacteria). This limitation can be removed by allowing the stress of the host to influence bacterial phenotypes (e.g. by subjecting $\beta_y$ to stress-dependent dynamics similar to that of $K_B$). This scenario is, in fact, expected in every holobiont due to the numerous options for 2-way interactions between the host and its symbionts. Such an influence could make a substantial contribution to the Lamarckian by enabling the emergence of new phenotypic values within one host generation. Since the current model restricts the emergence of new bacterial variations to mutations which occur on a characteristic timescale of $\mu^{-1}$ (1000 bacterial generations = 10 host generations), the Lamarckian adaptation is based only on selection of existing (bacterial) variants. Allowing the stress of the host to influence the bacterial phenotypes can support a stress-dependent increase in <$y_B$> due to changes during the lifetime of the host. Inheritance of these bacteria can then enhance the Lamarckian by taking advantage of new variations that emerge during the lifetime of the host.

The Lamarckian was evaluated by reverting a subset of holobionts to their initial state and re-subjecting them to toxin. Since this procedure cannot be realized experimentally, the Lamarckian of real holobionts should be approximated by other context-dependent means. For example, when symbiotic bacteria can be removed without compromising the survival of their toxin-free hosts (e.g. flies and worms that develop from dechorionated eggs on good diet conditions), the Lamarckian can be approximated in steps that are conceptually similar to the simulation procedure. First, the holobionts are exposed to a challenge and their offspring are then cleared of bacteria and separated into two subpopulations. One of these subpopulations is re-colonized with ('naïve') microbiota from untreated hosts (as in refs. (*17, 20*)), while the other is colonized by ('experienced') microbiota from a group of hosts which survived exposure to a challenge. The populations of hosts with naïve and experienced microbiota are then compared with respect to their ability to survive the challenge and the Lamarckian is approximated from the respective survival rates, $L \approx SR_{Exp.\ bact} / SR_{Naïve\ bact}) - 1$. This approximation, however, neglects other types of changes that may have been acquired and transmitted to offspring (e.g. small RNAs (*10*), maternal RNA (*39*), persistent chromatin modifications (*8*), horizontal transfer of biochemical signals (*40*) or other modes of local niche construction (*14*), etc). An additional source of deviation arises when the transmitted change is picked up by bystander offspring, which did not inherit the acquired change from their own parents. The extent to which these additional modes influence the evaluation of the Lamarckian may be assessed with a different ordering of the steps in the above procedure. The modified ordering starts with removal of bacteria from two untreated subpopulations and continues with re-colonizing these subpopulations with 'naïve' and 'experienced' microbiota, respectively, exposing the colonized subpopulations to the challenging condition and re-estimating the Lamarckian from their differential survival. More generally, it should also be possible to obtain a relative measure of the Lamarckian by manipulating the microbiome (or any other factor) in a subpopulation of hosts and evaluating the relative difference in offspring adaptation compared to offspring of non-manipulated parents (taken from the same distribution of hosts).

The above modelling framework does not aim to describe a particular holobiont, but rather to provide a general paradigm for considering adaptation to stress in a system with interactions between two levels of selection, each occurring on a different timescale. Combining this structure with pure Darwinian selection and constant rate of mutations gives rise to Lamarckian adaptation, group selection and stress-induced variability. These outcomes are not pre-specified but rather depend on the strength of toxic and physiological stress. Since the model relies on properties that are shared by diverse types of holobionts, the above conclusions are expected to be broadly applicable to a variety of holobionts and may also apply to other forms of equivalent organizations. Additionally, the modelling framework can be readily adjusted to consider additional factors, such as having multiple species of symbionts and pathogens (with inter-species competition and/or cooperation), asynchronous reproduction modes, epigenetic effects, ecological influences and transfer of bacteria (and/or toxin) between hosts.

**Acknowledgements**

We would like to thank Prof. Naama Brenner (Technion, Israel), Prof. Michael Shapira (UC Berkeley). Dr. Michael Elgart (Weizmann Inst. Israel), Maor Knafo (Weizmann Inst. Israel), Prof. Erez Braun (Technion, Israel) and Prof. David Bensimon (Paris ENS-LPS, France) for helpful discussions and suggestions. YR was supported by the Israel Science Foundation grant 1902/12 and the I-CORE Program of the Planning and Budgeting Committee of the Israel Council for Higher Education. DK was supported by the Binational Science Foundation, grant 2015619.

**References**

1. E. V. Koonin, Y. I. Wolf, Is evolution Darwinian or/and Lamarckian? *Biology direct* **4**, 42 (2009).
2. D. Noble, E. Jablonka, M. J. Joyner, G. B. Muller, S. W. Omholt, Evolution evolves: physiology returns to centre stage. *The Journal of physiology* **592**, (2014).
3. E. Rosenberg, G. Sharon, I. Zilber-Rosenberg, The hologenome theory of evolution contains Lamarckian aspects within a Darwinian framework. *Environmental microbiology* **11**, (2009).
4. K. Laland *et al.*, in *Nature*. (2014).
5. Y. Soen, Environmental disruption of host-microbe co-adaptation as a potential driving force in evolution. *Frontiers in genetics* **5**, (2014).
6. Y. Soen, M. Knafo, M. Elgart, A principle of organization which facilitates broad Lamarckian-like adaptations by improvisation. *Biology direct* **10**, 1-17 (2015).
7. E. Braun, The unforeseen challenge: from genotype-to-phenotype in cell populations. *Rep Prog Phys* **78**, (2015).
8. K.-H. Seong, D. Li, H. Shimizu, R. Nakamura, S. Ishii, Inheritance of stress-induced, ATF-2-dependent epigenetic change. *Cell* **145**, 1049-1061 (2011).
9. S. Stern, Y. Fridmann-Sirkis, E. Braun, Y. Soen, Epigenetically heritable alteration of fly development in response to toxic challenge. *Cell Rep* **1**, (2012).
10. O. Rechavi, G. Minevich, O. Hobert, Transgenerational inheritance of an acquired small RNA-based antiviral response in C. elegans. *Cell* **147**, (2011).
11. E. Jablonka, G. Raz, Transgenerational epigenetic inheritance: prevalence, mechanisms, and implications for the study of heredity and evolution. *The Quarterly review of biology* **84**, (2009).
12. E. V. Koonin, Viruses and mobile elements as drivers of evolutionary transitions. *Philosophical transactions of the Royal Society of London Series B, Biological sciences* **371**, (2016).
13. E. V. Koonin, Y. I. Wolf, Just how Lamarckian is CRISPR-Cas immunity: the continuum of evolvability mechanisms. *Biology direct* **11**, 9 (2016).
14. J. Odling-Smee, Niche Inheritance: A Possible Basis for Classifying Multiple Inheritance Systems in Evolution. *Biological Theory* **2**, 276–289 (2007).
15. I. Zilber-Rosenberg, E. Rosenberg, Role of microorganisms in the evolution of animals and plants: the hologenome theory of evolution. *FEMS microbiology reviews* **32**, (2008).

16. S. F. Gilbert, J. Sapp, A. I. Tauber, A symbiotic view of life: We have never been individuals. *The Quarterly review of biology* **87**, (2012).
17. Y. Fridmann-Sirkis *et al.*, Delayed development induced by toxicity to the host can be inherited by a bacterial-dependent, transgenerational effect. *Frontiers in genetics* **5**, 27 (2014).
18. E. Rosenberg, I. Zilber-Rosenberg, Symbiosis and development: the hologenome concept. *Birth Defects Res C Embryo Today* **93**, (2011).
19. M. Bakula, The persistence of a microbial flora during postembryogenesis of Drosophila melanogaster. *J Invertebr Pathol* **14**, 365-374 (1969).
20. M. Elgart *et al.*, Impact of gut microbiota on the fly's germ line. *Nature communications* **7**, 11280 (2016).
21. L. Margulis, Ed., *Symbiosis as a source of evolutionary innovation: speciation and morphogenesis*, (Cambridge: MIT press, 1991), pp. 1-14.
22. R. A. Fisher, The wave of advance of advantageous genes. *Ann. Eugenics* **7**, 353-369 (1937).
23. A. J. Lotka, Analytical Note on Certain Rhythmic Relations in Organic Systems. *Proc. Nat'l. Acad. Sci. (USA)* **6**, 410-415 (1920).
24. V. Volterra, Variazioni e fluttuazioni del numero d'individui in specie animali conviventi. *Mem. Acad. Lincei Roma* **2**, 31-113 (1926).
25. F. B. a. C. Castillo-Chavez, *Mathematical Models in Population Biology and Epidemiology*. (Springer-Verlag, New York, 2000).
26. W. D. H. a. R. M. May, Dispersal in Stable Habitats. *Nature* **269**, 578-581 (1977).
27. M. A. Nowak, *Evolutionary Dynamics: Exploring the Equations of Life*. (Belknap Press, Cambridge, MA, 2006).
28. P. L. Antonellia, S. F. Rutzb, P. W. Sammarcoc, K. B. Strychard, Evolution of symbiosis in hermatypic corals: A model of the past, present, and future. *Nonlinear Analysis: Real World Applications* **32**, 389–402 (2016).
29. R. van Woesik, K. Shiroma, S. Koksal, Phenotypic variance predicts symbiont population densities in corals: a modeling approach. *PloS one* **5**, e9185 (2010).
30. I. Sekirov, S. L. Russell, L. C. Antunes, B. B. Finlay, Gut microbiota in health and disease. *Physiol Rev* **90**, 859-904 (2010).
31. L. Mouton, H. Henri, D. Charif, M. Bouletreau, F. Vavre, Interaction between host genotype and environmental conditions affects bacterial density in Wolbachia symbiosis. *Biol Lett* **3**, 210-213 (2007).
32. L. Muscatine, R. R. Pool, Regulation of numbers of intracellular algae. *Proc R Soc Lond B Biol Sci* **204**, 131-139 (1979).
33. Y. Fishman, E. Zlotkin, D. Sher, Expulsion of symbiotic algae during feeding by the green hydra--a mechanism for regulating symbiont density? *PloS one* **3**, e2603 (2008).
34. S. D. Siciliano *et al.*, Selection of specific endophytic bacterial genotypes by plants in response to soil contamination. *Applied and environmental microbiology* **67**, 2469-2475 (2001).
35. J. A. Ceja-Navarro *et al.*, Gut microbiota mediate caffeine detoxification in the primary insect pest of coffee. *Nature communications* **6**, 7618 (2015).
36. C. Tetard-Jones, R. Edwards, Potential roles for microbial endophytes in herbicide tolerance in plants. *Pest Manag Sci* **72**, 203-209 (2016).
37. C. H. Waddington, GENETIC ASSIMILATION OF THE BITHORAX PHENOTYPE. *Evolution* **10**, 1-13 (1956).

38. G. Gibson, D. S. Hogness, Effect of polymorphism in the Drosophila regulatory gene Ultrabithorax on homeotic stability. *Science* **271**, 200-203 (1996).
39. S. Stern *et al.*, Reduction in maternal polycomb levels contributes to transgenerational inheritance of a response to toxic stress in flies. *The Journal of physiology* **592**, 2343-2355 (2014).
40. G. A. Miller, M. S. Islam, T. D. Claridge, T. Dodgson, S. J. Simpson, Swarm formation in the desert locust Schistocerca gregaria: isolation and NMR analysis of the primary maternal gregarizing agent. *J Exp Biol* **211**, 370-376 (2008).

**Supplementary Methods**:

The simulation starts with a population of hosts, each carrying a population of 100 bacteria. Host and bacterial properties (phenotypes) are initially drawn from defined distributions (steady state of Eq. 5 without toxin) with the parameters $x_0$ = 0.25, $\beta_x$ = 10, $y_0$ = 0, $\beta_y$ = 0.1, $\delta_0$ = 0 and $\beta_\delta$ = 0.1.

In every time step of the simulation (one bacterial generation), each bacterium reproduces if its survival probability (Eq. 2) is larger than a random number (between 0 and 1) drawn from a uniform distribution. Each of the surviving bacterial (parent) persists at its current state and gives rise to a modified bacterium (offspring), while dead bacteria are discarded. At the end of one host generation (100 time steps), the reproduction of hosts is determined based on the survival probability in Eq. 1. Non-surviving hosts are discarded and each of the surviving hosts gives rise to a parent and offspring host as follows:

- The parent retains its current state (sensitivity, detox, delta) and the state of its bacterial population.
- Following 99 bacterial generations, an offspring host is created with properties defined by Eq. 5. Negative values of the sensitivity and detox are prevented by taking the absolute value of the outcome in Eq.1. Each offspring receives a copy of the bacterial population of its parent. These populations are then iterated forward one bacterial generation, the surviving bacteria reproduce so as to define the initial state of the bacterial populations in the next host generation of the parent and its offspring.

**Supplementary Figures:**

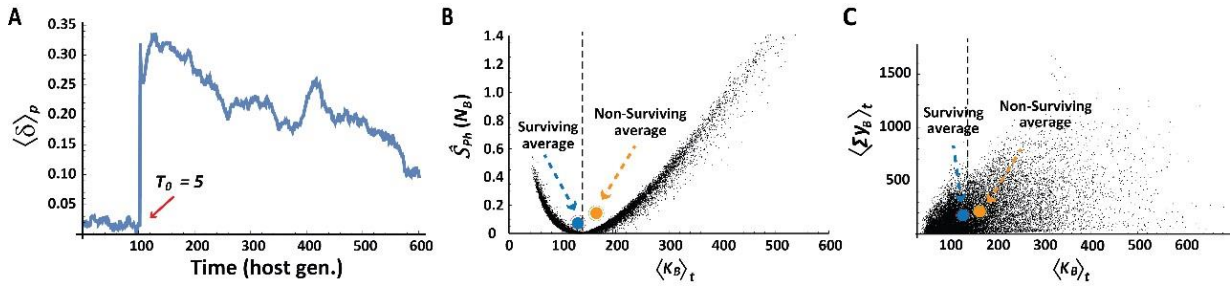

**Figure S1: (A)** Rapid selection of hosts with large δ under exposure to a pulse of toxin ($T_0$=5), applied at the initial time step, $t \in [0, \Delta t]$ (red arrow). **(B)** Host physiological stress over a host generation, $\hat{S}_N$ = $ln(<N_B>_t/K_B^0) + (1 - <N_B>_t/K_B^0)$, versus time average of $K_B$ (Same as Fig. 1E, except for the exclusion of hosts which lost their bacteria). Blue and orange circles mark population averages for surviving and non-surviving holobionts, respectively. Dotted line marks the $K_B$ value which minimizes the physiological stress. **(C)** Same as (B) for the time average of total bacterial detox versus bacterial carrying capacity.

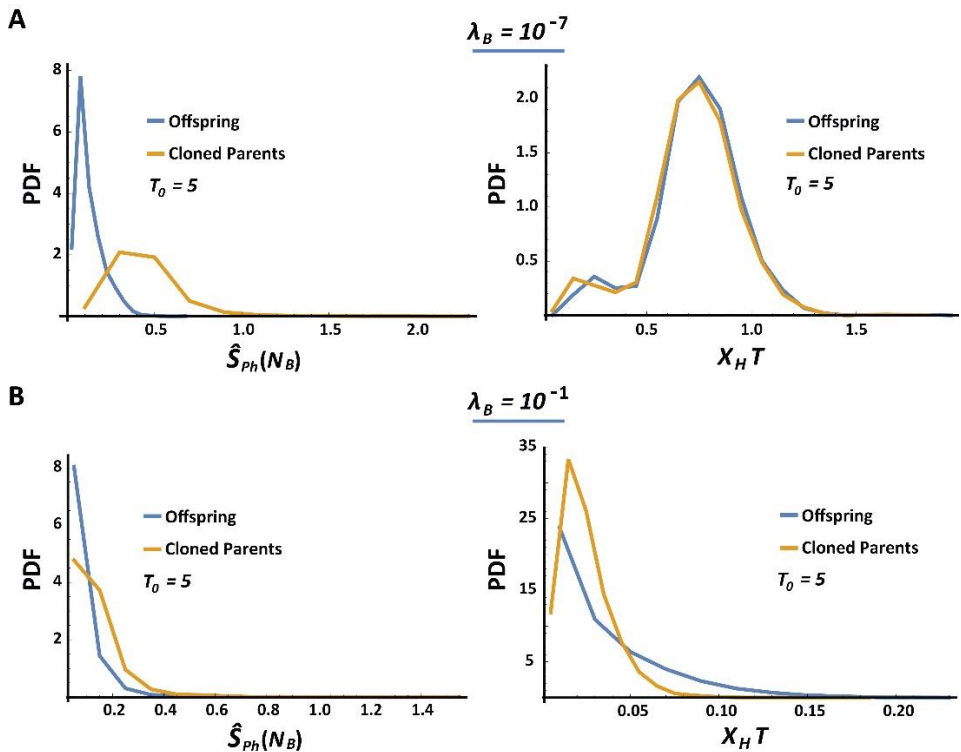

**Figure S2:** Distributions of instantaneous physiologic (left) and toxic stress (right) in cloned parents and their offspring, immediately after one generation of exposure to toxin ($T_0$=5). **(A)** Case of low detox coefficient, $\lambda_B$=$10^{-7}$. **(B)** Case of high detox coefficient, $\lambda_B$=$10^{-1}$.

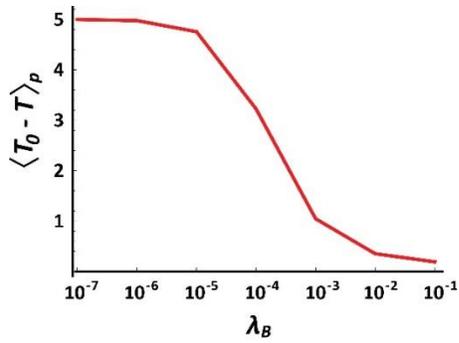

**Figure S3:** Average level of active toxin at the end of one host generation as a function of bacterial detox coefficient.

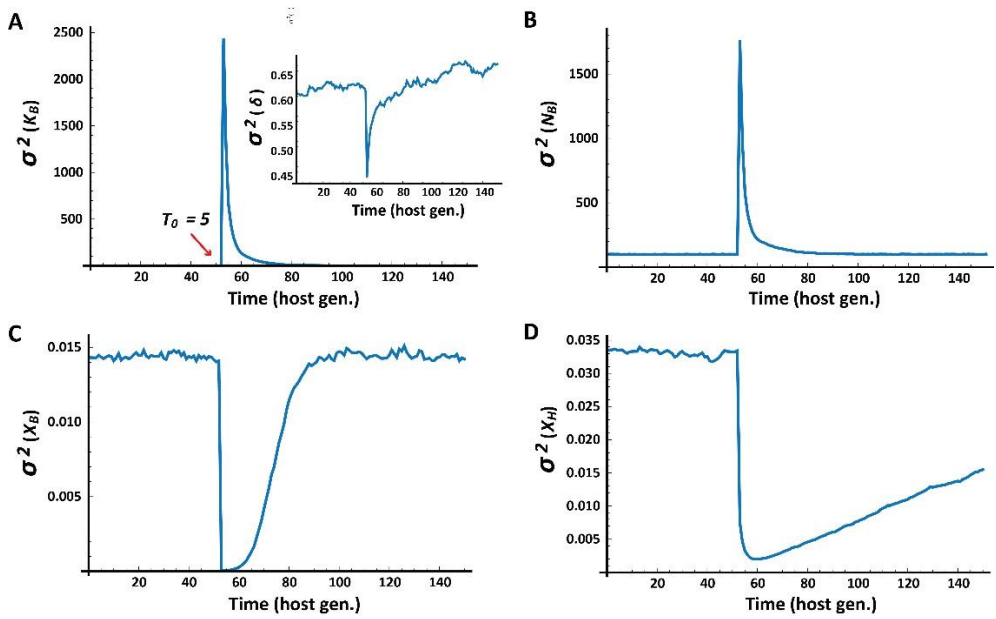

**Figure S4: Temporal kinetics of phenotypic variability in response to toxic exposure.** The holobiont population is exposed to a toxin pulse ($T_0$ =5) at generation 50 (red arrow in A). Shown are instantaneous inter-holobiont variability in the bacterial carrying capacity (A) and host $\delta$ (A, inset), size of the bacterial population (B), average bacterial sensitivity in the holobiont (C) and average host sensitivity (D).